\documentclass[aps, prfluids,twocolumn,notitlepage,superscriptaddress,showpacs,10pt]{revtex4-2}

\usepackage{hyperref}
\usepackage{graphicx}
\usepackage{dcolumn,xcolor}
\usepackage{bm}
\usepackage{soul}
\usepackage{amsmath}
\begin{document}

\preprint{APS/123-QED}

\title{Droplet coalescence in fluids obeying Darcy's law}

\author{Jing Wang}
\affiliation{Department of Physics, Emory University, Atlanta, GA 30322, USA.}
\author{Haicen Yue}
\affiliation{Department of Physics, University of Vermont, Burlington, VT 05405, USA.}
\author{Nandish Vora}
\affiliation{Department of Physics and Astronomy, University of Pennsylvania, Philadelphia, PA 19104, USA.}
\author{Tabitha C. Watson}
\affiliation{Department of Mechanical Engineering, University of Michigan, Ann Arbor, MI 48109, USA.}
\author{Zhengyan Lin}
\affiliation{School of Physics, Georgia Institute of Technology, Atlanta, GA 30332, USA.}
\author{Itamar Kolvin}
\affiliation{School of Physics, Georgia Institute of Technology, Atlanta, GA 30332, USA.}
\author{Justin C. Burton}%
\email{justin.c.burton@emory.edu}
\affiliation{Department of Physics, Emory University, Atlanta, GA 30322, USA.}

\date{\today}

\begin{abstract}
During drop coalescence, a connecting bridge of fluid forms and rapidly expands due to surface tension. For spherical drops, these dynamics are well understood in both the viscous and inertial regimes. However, under strong confinement, fluid motion is fundamentally altered by geometric constraints, leading to dissipation on small lengthscales. We investigate the coalescence of drops confined in a Hele-Shaw cell (two parallel plates separated by a narrow gap). In this geometry, the depth-averaged flow is governed by Darcy’s law while surface tension drives the interface motion. We identify two distinct temporal regimes in the evolution of the bridge radius that evolves as a power law ($R_b$). At early times, the bridge grows as $R_b \sim t^{1/2}$, which results from a confinement-dependent meniscus instability that determines the initiation of contact between droplets prior to bridge formation. At later times, the bridge growth slows substantially and follows $R_b \sim t^{1/5}$, consistent with recent theoretical predictions for Darcy-governed coalescence. We show that the transition between these regimes is controlled by several geometric lengthscales. In particular, the onset of the Darcy regime occurs when the interface radius of curvature becomes comparable to the plate spacing, such that the flow becomes fully confined. Using a boundary integral formulation, we find that both scaling laws for $R_b$ are determined by the bridge width. Together, these results identify a new universal regime of drop coalescence in a broad class of fluids obeying Darcy's law.

\end{abstract}

\maketitle


\section{Introduction}

The coalescence of fluid droplets is a classical hydrodynamic problem in which surface tension drives topological rearrangement of an interface through a singularity. Coalescence plays a central role in a wide range of industrial and natural processes, including inkjet printing \cite{hong2022coalescence, jones1998characterizing}, pharmaceuticals \cite{haghighat2020droplet, haghighat2019droplet}, oil spill clean-up \cite{adofo2022dispersants, sun2009review}, and cell aggregates \cite{oriola2022arrested, thomas2014dynamics}. Understanding the theoretical framework governing coalescence is therefore essential for controlling industrial manufacturing processes and interpreting natural phenomena. At its core, coalescence converts surface energy into fluid motion, which is ultimately dissipated by viscosity. These dynamics are concentrated around the small bridge region that connects the drops, and are affected by fluid properties such as viscosity and inertia, as well as external constraints \cite{eggers2025coalescence}. In particular, geometric confinement can lead to a new class of coalescence dynamics that differs fundamentally from unconfined fluid systems \cite{yokota2011dimensional,yue2024coalescing}. 

The theoretical framework of viscous drop coalescence was initiated by Frenkel using a balance of surface tension and viscous forces during the sintering process \cite{frenkel1945viscous}. The analytical solution of two coalescing viscous cylinders was later developed by Hopper, which also characterizes the initial shape evolution and dynamics of spherical drops \cite{hopper1993coalescence1, hopper1993coalescence2}. Hopper's solution has been verified repeatedly by experimental and numerical studies in the following years \cite{eggers1999coalescence,paulsen2013approach, wu2004scaling, anthony2020initial, sprittles2014parametric, paulsen2012inexorable}. 
Hopper's solution is valid for purely viscous fluids, where momentum diffuses much faster than the rate of bridge expansion. Using an electrical method to measure the bridge dynamics, combined with simulations, \citet{paulsen2011viscous,paulsen2012inexorable}  showed that the finite inertia of the drops can affect the initial moments of coalescence. Despite their differences, these above classical descriptions all assume hydrodynamic transport governed by momentum-conserving viscous flow. Many experimentally relevant systems, however, have system-specific features such as viscoelasticity, non-reciprocity, porous flow, and geometrical bounds that cannot be purely described by Stokes flow. A notable recent study by \citet{yue2024coalescing} demonstrated that coalescence in momentum-dissipative systems can be described by Darcy's law, which should universally apply to both highly confined drops and also particulate and cellular matter coupled to an external bath. However, coalescence dynamics of fluids obeying Darcy's law remain a mystery without a corresponding analytical theory.





\begin{figure*}[t]
\centering
\includegraphics[width=1\textwidth]{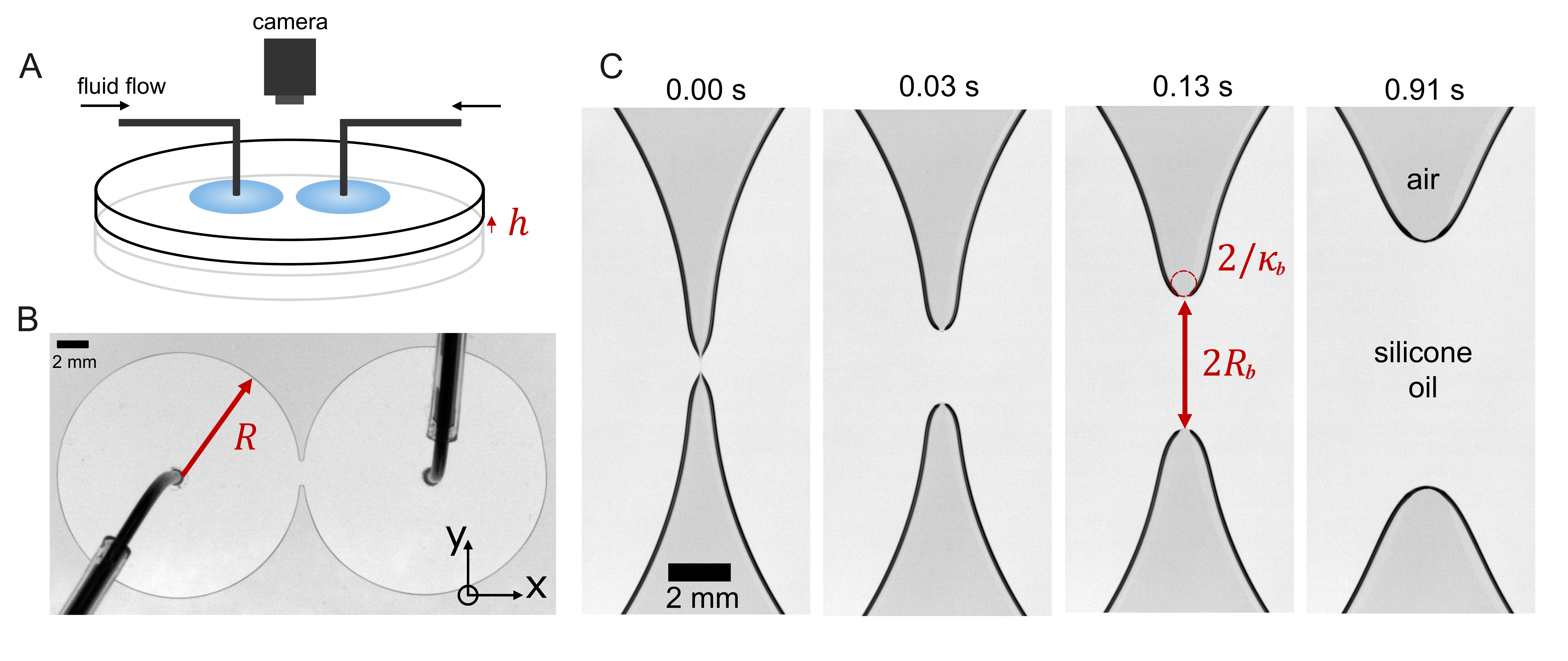}
\caption{Observing droplet coalescence in a quasi-2D Hele-Shaw cell. (A) Silicone oil is injected between two glass disks that are separated by a gap size $h$, which is determined by spacers that are hundreds of microns in thickness. The fluid is split between two tubes and pumped at the same flow rate, producing two equal-sized pancake drops. A camera is mounted from above to capture the coalescence process. (B) A sample image of two 97 mPa$\cdot$s silicone oil drops coalescing captured from above. A video of the coalescence process is shown in Movie \textbf{S1}. (C) Zoom-in of the coalescence dynamics. The drop interfaces deform and initiate coalescence at a small finite separation. A zoomed-in video of coalescence captured by a macro lens is shown in Movie \textbf{S2}.}
\label{fig:experiment setup}
\end{figure*}

One canonical example of momentum-dissipative flow arises in Hele-Shaw geometries, where fluids are confined in a narrow gap between two solid surfaces, and the depth-averaged flow is governed by Darcy’s law. This geometry has predominantly been used to study viscous fingering \cite{liu2026effect,saffman1986viscous,homsy1987viscous}, yet has recently been used to investigate drop coalescence under confinement, with most studies focusing on the dynamics of bridge growth \cite{yokota2011dimensional,ryu2022dark, eri2010bursting,dolganov2021dynamics, delabre2010coalescence, shuravin2019coalescence}. 
However, in nearly all of these studies, the gap thickness $h$ remained comparable to the drop radius ($R$), or the bridge radius ($R_b$) during coalescence, meaning the coalescence geometry retained a three-dimensional structure. As a result, the corresponding hydrodynamics did not fully reach the Darcy limit expected for strongly confined, pancake-shaped droplets. In this limit, the confinement scale $h$ must be smaller than any other lengthscale.

Here we investigate droplet coalescence in a Hele-Shaw geometry, where strong confinement introduces viscous friction and momentum dissipation. Throughout the coalescence process, we ensure that $h \ll R$, yet we still observe two dynamical regimes. At early times, the radius of curvature of the connecting bridge between the drops is unavoidably smaller than $h$, meaning the coalescence is not purely described by Darcy's law, and the bridge dynamics follow $R_b\sim t^{1/2}$. At later times, we observe bridge growth dynamics consistent with Darcy's law, where $R_b \sim t^{1/5}$ \cite{yue2024coalescing}. The initial contact between the drops is determined by a short-ranged attraction between the confined menisci at the edge of the drops, leading to a ``jump to contact.'' Using Particle Image Velocimetry (PIV), we show that the flow and dissipation are concentrated around the interfaces near $R_b$, rather than distributed throughout the entire connecting bridge. We use these experimental observations in conjunction with an energy balance argument to explain the bridge growth scaling laws. Furthermore, we develop a boundary integral analysis of the flow and identify the relevant lengthscales that define coalescence in the Darcy regime. More broadly, our results demonstrate that drop coalescence remains a rich subject in fluid mechanics where the dynamics are sensitive to the geometry of the singular connecting bridge between the drops. 

\section{Experimental Methods}

A custom Hele-Shaw cell was constructed using two circular glass plates of radius 50 mm and thickness 6.25 mm. The gap between the plates ($h$) was  controlled using thin polyester shims (Fig.~\ref{fig:experiment setup}A), so that $h$ varied between 127-762 $\mu$m. For all experiments, the droplet radius was much larger than the gap thickness ($h/R \ll 1$), resulting in strongly confined, pancake-shaped droplets. During an experiment, viscous silicone oil (Clearco) was simultaneously injected through two metal tubes connected to holes drilled in the glass plates. The density of the oil was 965 kg/$m^3$ $\pm$ 5 kg/$m^3$, and the viscosity ranged from 48 to 971 mPa$\cdot$s. The tube diameter was 1.6 mm, and the tubes were separated by 30 mm, as shown in Fig.~\ref{fig:experiment setup}B. We used a syringe pump (NE-1000, New Era) to precisely control the growth rate of the pancake-shaped drops confined within the thin gap. The pumping speed was set to 5 mL/hr to  avoid residual growth of the drops when the pumping ceased (i.e., from elastic expansion of the Tygon tubing).

To initiate coalescence, fluid was injected until moments before the drops contacted, at which point the syringe pump was manually stopped. Coalescence started within 1-3 s after stopping the pump. The subsequent coalescence dynamics therefore occurred in the absence of continued fluid injection. We used a macro lens connected to a digital camera (Point Grey) to resolve the bridge region during coalescence. For most experiments, the fluid interface was recorded from above at 90 frames per second with a spatial resolution of 11.5 $\mu$m/pixel (Fig.~\ref{fig:experiment setup}C). 
The fluid-air interface was extracted from binarized image sequences via thresholding and image analysis. The bridge radius $R_b$ was obtained by measuring the vertical distance between the opposing interfaces at the location of the liquid bridge. The curvature $\kappa_{b}$ was measured through fitting the local areas of the minimum and maximum region to a 4th-order polynomial and subsequently calculated via its derivatives evaluated at $R_b$. Each dataset of $R_b$ and $\kappa_b$ contained approximately 13,000 measurements. For clarity in visualization, the data were logarithmically binned into $\sim30$ points for plotting.

To measure flow fields during coalescence, we employed Particle Image Velocimetry (PIV) using the open-source MATLAB package PIVlab \cite{thielicke2014pivlab}. We chose commercial cocoa powder as neutrally buoyant tracer particles, which produced highly visible traceable particles and did not sediment throughout the coalescence experiment. Prior to PIV analysis, the image contrast was enhanced in ImageJ, and masks defining the fluid-air interface were applied to exclude spurious vectors outside the droplet region. Velocity fields were computed using a multi-pass Fast Fourier Transform window-deformation algorithm with progressively decreasing interrogation window sizes. To account for the rapidly decreasing flow velocity during coalescence, frames were sampled with logarithmically increasing temporal spacing so that the displacement between frames was approximately constant.

\section{Hele-Shaw Flow and Darcy's Law}
\label{sec:helederiv}

The governing equation for flow in a Hele-Shaw geometry is Darcy's law, which is valid under the assumption that the gap size $h$ is much smaller than the variations in flow in the $x-y$ plane. Darcy's law can be directly derived from the Stokes equation for incompressible flow: 
\begin{equation}
    \bm{\nabla} P - \eta \nabla ^2 {\bm{v}} = 0. \label{darcy der: stokes}
\end{equation}
Here, $\bm{\nabla} P$ is the pressure gradient, $\eta$ is the dynamic viscosity of the fluid, and $\bm{v}$ is the flow velocity. We assume a viscous, incompressible flow between the plates with no slip on the boundaries, where the flow profile is parabolic in the $z$ direction, so that the total velocity is 
\begin{equation}
    {\bm v} = A(h-z)z\left[ v_x(x, y), v_y(x, y),0 \right]. \label{darcy der: generic v}
\end{equation}
The depth-averaged velocity takes the form
\begin{equation}
    \langle{\bm v}\rangle = \frac{1}{h}\int^h_0 {\bm v} dz=\left[ v_x(x, y), v_y(x, y),0 \right]. \label{darcy der: generic <v>}
\end{equation}
 We can directly solve for the prefactor $A$ by substituting Eq.~\ref{darcy der: generic v} into Eq.~\ref{darcy der: generic <v>} and evaluating the integral. As a result, 
\begin{equation}
    {\bm v} = \frac{6}{h^2}z(h-z)\left[ v_x(x, y), v_y(x, y),0 \right], \label{darcy der: v}
\end{equation}
where $v_x$ and $v_y$ are the depth-averaged horizontal velocities. Plugging the full expression of ${\bm v}$ into the Stokes equation (Eq. \ref{darcy der: stokes}), with the assumptions
\begin{equation}
    \frac{v_x}{h^2} \gg \nabla^2v_x,\hspace{12pt}\frac{v_y}{h^2} \gg \nabla^2v_y, \label{velgrad}
\end{equation}
we arrive at Darcy's law for the depth-averaged horizontal flow in the gap:
\begin{equation}
    -\frac{h^2}{12 \eta} \bm{\nabla} P = \langle{\bm{v}}\rangle. \label{darcy der: darcy final boss}
\end{equation}

The key assumption is that gradients in velocity are dominated by gradients in $z$. This is only true if the interface geometry varies over lengthscales larger than $h$. Since drop coalescence contains a singularity at the moment of contact, the interface curvature is very large at early times, and undoubtedly $1/\kappa_{b}\sim h$, violating the assumptions used to derive Darcy's law. 
Thus, the assumption that the gap size is sufficiently small is a deciding factor of whether Darcy's law applies in the Hele-Shaw geometry. 

The gap size is also a limiting factor when determining the Laplace pressure across the interface. During coalescence, surface tension ($\gamma$) determines the pressure jump between the fluid and surrounding air. There are two primary curvatures. The first is the in-plane curvature $\kappa_{b}$ given by the shape of the fluid-air interface when viewed from above, as in Fig.~\ref{fig:experiment setup}C. However, the interface is really a meniscus between the glass plates. Since silicone oil wets the glass surface (zero contact angle), we can locally approximate this meniscus shape as half of a circle with diameter $h$. Thus, the total curvature is given by $\kappa_{b}+2/h$. Typically in Hele-Shaw flow, the larger curvature (2/$h$) is considered constant along the interface and treated as a constant pressure jump that can be ignored. Thus only the in-plane curvature varies along the interface and can drive fluid flows. However, when $1/\kappa_{b}\sim h$, the interface shape is really three-dimensional, and the total curvature may not be partitioned cleanly. 

Throughout this work, we will directly compare our experimental results to a computational solution of two-dimensional coalescence of drops that obey Darcy's law. The solution follows a boundary integral formulation first presented in \citet{yue2024coalescing}, and full details are included in the appendix. 
Briefly, we can nondimensionalize Eq.~\ref{darcy der: darcy final boss} using the following length, time, and pressure scales: $R$, $\tau$, and $\gamma/R$, where the timescale is given by:
\begin{equation}
    \tau = \frac{12 \eta R^3}{\gamma h^2}. \label{taueq}
\end{equation}
Then Eq.~\ref{darcy der: darcy final boss} becomes:
\begin{equation}
    \tilde{\nabla} \tilde{P} = - \tilde{\bm{v}},
\end{equation}
where the tilde indicates dimensionless variables, and we have dropped the angle brackets on the velocity for brevity. It is immediately clear that coalescence follows potential flow, and the pressure is the velocity potential. Taking the divergence of both sides, and assuming incompressible flow, we reduce the problem to solving a Laplace equation for the pressure with a boundary condition that depends on the local curvature:
\begin{align}
    \tilde{\nabla}^2 \tilde{P} &= 0,\\
    \tilde{P}_1 - \tilde{P}_2 &= \tilde{\kappa}.
    \label{eq:bcond}
\end{align}
The subscripts designate inside (1) the coalescing drops, and outside (2). This is the same formulation needed to solve coalescence in the Stokes regime \cite{eggers1999coalescence}, but the relationship between the velocity and pressure is given by Darcy's law, not the Stokes equation for creeping flow. For two fluids, Eq.~\ref{eq:bcond} plus the requirement that the normal velocity be continuous across the boundary is enough to generate a unique solution to the harmonic functions $P_1$ and $P_2$. If there is only an inner fluid, then the pressure $P_1$ at the boundary is just defined by the curvature, and the normal velocity is predetermined. 

Following Pozrikidis \cite{pozrikidis1992boundary}, we used the double layer boundary integral formulation to find the velocity at any point on the surface, given the boundary shape. For simplicity and to avoid solving an integral equation, we assume that the outer and inner fluids have equal viscosities. In this limit, the velocity of the boundary is given by:
\begin{equation}
    {{\bm v}}(x_0) = \oint \bm{\nabla} G({\bm x}_0,{\bm x})\times\left[\bm{\nabla}\kappa({\bm x}) \times \bm n({\bm x})\right] ds(\bm x), \label{bintvel}
\end{equation}
where we have dropped the tilde over the dimensionless variables for simplicity. $G({\bm x}_0,{\bm x})$ is the two-dimensional Green's function for the Laplace equation, and the unit vector normal to the interface is $\bm{n}$. Coalescence proceeds by using an initial shape (i.e., from the analytical Hopper solution \cite{hopper1993coalescence1,hopper1993coalescence2}), computing the velocity, advancing the interface, and repeating. Full details are included in Appendix \ref{appendA}. Equation~\ref{bintvel} will additionally be used in Sec. VII to derive scaling relations for the bridge growth dynamics.


\section{Bridge Dynamics and Geometry}

\begin{figure}[!]
\centering
\includegraphics[width=\columnwidth]{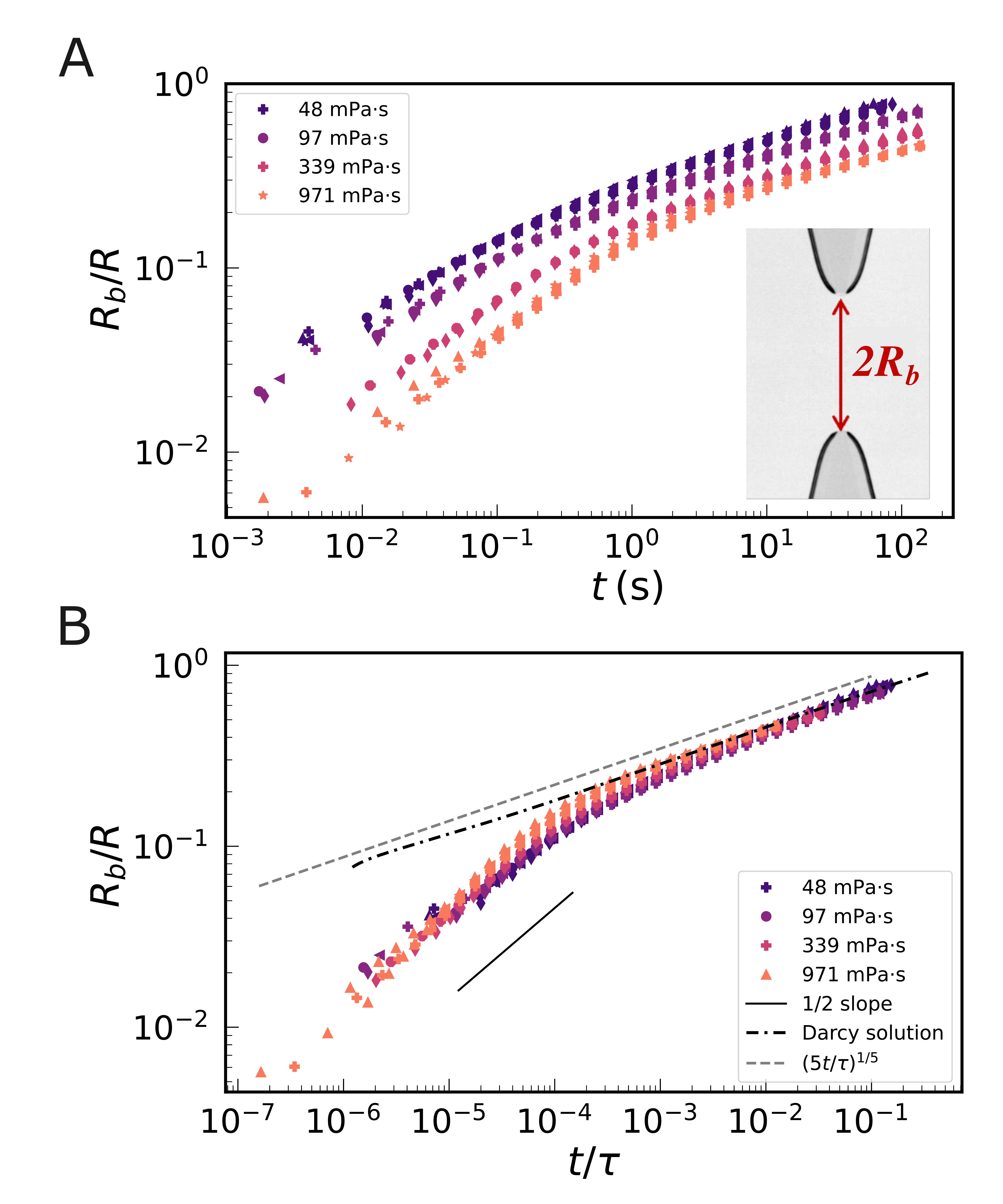}
\caption{(A) Dynamics of the normalized bridge radius $R_b/R$ show two distinct regimes. Data are shown for silicone oil with various viscosities at a gap size of $h$ = 400 $\mu$m. We show 3-4 datasets for each viscosity, demonstrating the high repeatability of the measurements. The inset shows the bridge radius $R_b$ defined in one sample frame. (B) The data can be collapsed using the timescale $\tau$ from Eq.~\ref{taueq}. The gray dashed line represents the scaling law given by Eq.~\ref{t 1/5 relation}, and the black dot-dashed line is the Darcy simulation provided by solving Eq.~\ref{bintvel}. The solid black line represents a slope of 1/2, as estimated by the data at early times.  }
\label{fig:neck 1}
\end{figure}

Figure \ref{fig:neck 1}A shows the time evolution of $R_b/R$ for silicone oil of various viscosities with gap size $h$ = 400 $\mu$m. Due to the finite resolution of the imaging system, there was uncertainty in the precise moment of initial contact. In Fig.~\ref{fig:neck 1}, we define $t = 0$ s as the moment of coalescence by fitting the first 20 data points of the raw data to a power-law functional form, $R_b(t') = a (t'-t_0)^b$. We subsequently shifted all data by $t_0$ to align all experiments. The bridge growth exhibits two distinct regimes with different scaling behaviors, where the transition occurs at $R_b/R\approx0.1-0.2$. This crossover naturally emerges because $h$ is comparable to the smallest lengthscales, for example, $1/\kappa_{b}\sim h$. For Fig.~\ref{fig:neck 1}, the gap size was fixed, so the speed of coalescence was mostly determined by the viscosity. The behavior of $R_b$ vs. $t$ is consistent with a power law in both regimes, where $R_b \propto t^\beta$.

To collapse the data and make this scaling apparent, we use the timescale $\tau$ from Eq.~\ref{taueq}. The dominant pressure gradient arises from the highly curved bridge region. We approximate this gradient as $|\bm{\nabla}P|\sim\gamma\kappa_{b}/L$, where $L$ is a lengthscale that defines how the pressure decays away from $R_b$ in the vertical direction (along the red arrow in Fig.~\ref{fig:neck 1}A). As a rough estimate, we assume that  $1/\kappa_{b} \sim L\sim R_b^2/R$. This choice comes from previous analysis of coalescence under Darcy's law using Brownian dynamics simulations of particle clusters and boundary integral simulations \cite{yue2024coalescing}. With this assumption, we obtain the following differential equation for the bridge radius $R_b$:
\begin{equation}
    \bm v\cdot\hat{\bm y}=\frac{dR_b}{dt} = \frac{\gamma R^2}{R_b(t)^4}\frac{h^2}{12 \eta}, \label{deriving 1/5}
\end{equation}
Solving this equation gives:
\begin{equation}
    R_b(t) = \bigg[ \frac{5 \gamma R^2 h^2}{12 \eta}t\bigg] ^{1/5}. \label{t 1/5 relation}
\end{equation}
which can be written as $R_b/R = (5t/\tau)^{1/5}$ using Eq.~\ref{taueq}. This is plotted as a gray dashed line in Fig.~\ref{fig:neck 1}. The power-law scaling agreement is reasonable and within a multiplicative factor of 2 from the data.  


Figure \ref{fig:neck 1}B shows the collapse of the data presented by normalizing time with $\tau$. The quality of the data collapse suggests that at fixed gap size, the timescale for dynamics in both regimes depends linearly on viscosity. We observe that $\beta$ is close to $1/2$ at early times and is close to $1/5$ at late times, as shown by the solid and dashed lines, respectively.  The solution of the boundary integral method \cite{yue2024coalescing} is also shown with no adjustable parameters. Our results are consistent with the Brownian dynamics simulations from \citet{yue2024coalescing}, where $\beta\sim0.4$ at early times, and crosses over to a much smaller scaling at later times where $\beta$ is approximately $1/5$. We note that the choice of $1/\kappa_{b}\sim L$ is an assumption at this point, and more details on the relation between these lengthscales will be provided in Sec.~VII. Furthermore, we note that the agreement between these disparate systems, i.e., Brownian particle dynamics and confined fluid flow, illustrates the universality of coalescence in systems obeying Darcy's law. 

\begin{figure}[t!]
\centering
\includegraphics[width=\columnwidth]{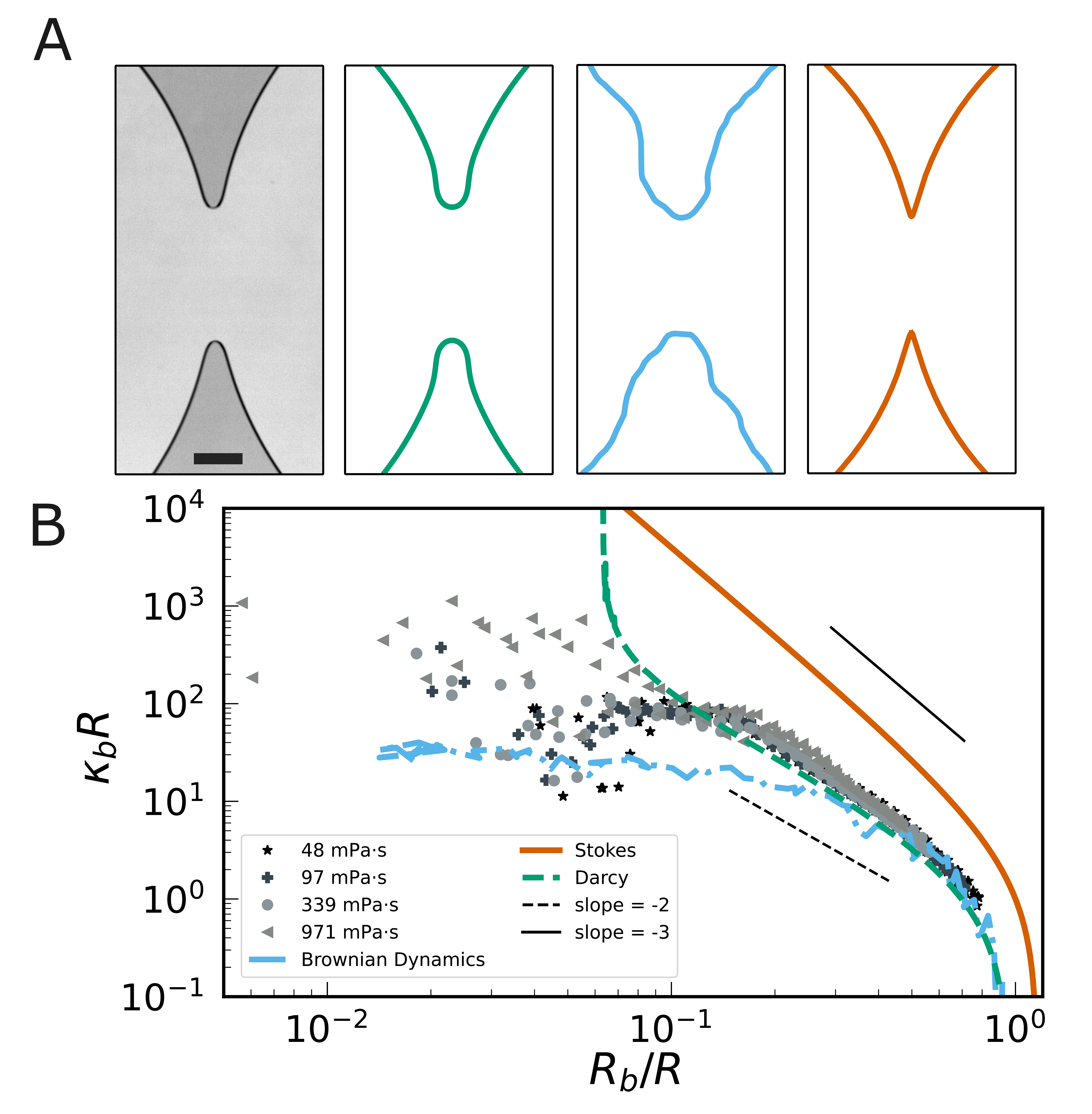}
\caption{ (A) A comparison of the bridge geometry at $R_b/R \approx 0.2$ for (from left to right): a Hele-Shaw cell experiment ($\eta$ = 97 mPa$\cdot$s, $h$ = 400 $\mu$m); Darcy coalescence simulation (Eq.~\ref{bintvel}); Brownian dynamics simulations \cite{yue2024coalescing}; Hopper solution for Stokes flow \cite{hopper1993coalescence1,hopper1993coalescence2}. The scale bar in the leftmost panel represents 2 mm. (B) The bridge curvature at the minimum, $\kappa_{b}$, normalized by the drop radius $R$, plotted as a function of the normalized bridge radius $R_b/R$. The data are the same experiments shown in Fig.~\ref{fig:neck 1}. The orange solid line represents the Hopper solution according to Stokes flow, the green dashed line represents the Darcy simulation, and the blue line represents the Brownian dynamics simulation. The black dashed and solid lines are guides to the eye with slopes of -2 and -3, respectively.}
\label{fig:contour compare and curvature}
\end{figure}

The early-time behavior where $\beta \approx 1/2$ is due to the finite gap size, i.e., when $h\sim 1/\kappa_b$. This sets the ultimate limit where Darcy's law can be applied. Before understanding why $\beta\approx 1/2$, we examine the geometry of the bridge curvature. Figure \ref{fig:contour compare and curvature}A shows bridge geometry when $R_b/R = 0.2$ for our experiments, the boundary integral solution of Darcy's law, the Brownian dynamics from \citet{yue2024coalescing}, and finally for the Hopper solution for Stokes flow coalescence \cite{hopper1993coalescence1,hopper1993coalescence2}. Compared to the experiment, the Darcy solution is less sharp with a smaller curvature at the bridge. This is presumably due to the presence of the outer fluid in the simulation, which can be trapped like a ``bubble,'' as described in other coalescing droplet systems \cite{eggers1999coalescence}. The Brownian dynamics simulations have a significantly smaller curvature and exhibit interface roughness due to the finite particle sizes used in the simulation. All three of these cases have much smaller curvatures than the analytical solution for viscous drops (Hopper solution). This stems from the translation of the center of mass of each drop as they approach each other \cite{paulsen2012inexorable,yue2024coalescing}.

\begin{figure}[t!]
\centering
\includegraphics[width=\columnwidth]{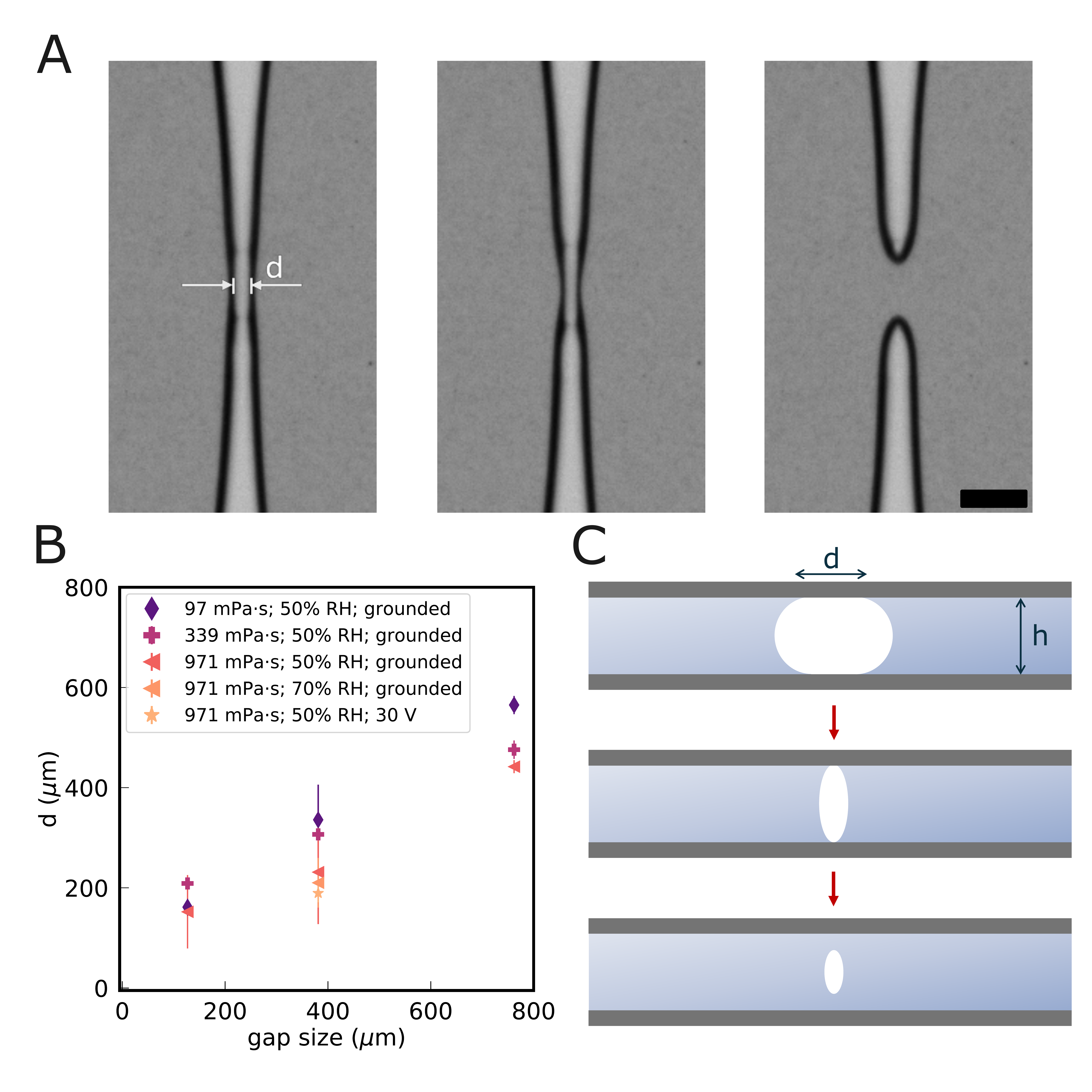}
\caption{Investigating the length scale $d$ at the moment of coalescence contact, defined as the minimum horizontal distance between two drops at the moment of contact. (A) Image sequences right before moment of coalescence contact at 200 fps for a 971 mPa$\cdot$s silicone oil at $h = 400$ $\mu$m. The meniscus is visible in the background of the fluid outline. Scale bar indicates 500 $\mu$m. A video of the contact formation is shown in Movie \textbf{S3}. (B) The finite approach length scale $d$ increases as the gap size increases. Varying the relative humidity (RH) and the electrostatics conditions do not change the overall dependence on $h$. (C) A hypothetical cartoon of the meniscus evolution at the moment of coalescence in the \textit{x-z} plane. }
\label{fig:meniscus}
\end{figure}

Figure \ref{fig:contour compare and curvature}B shows the normalized curvature, $\kappa_b R$, as a function of normalized bridge radius, $R_b/R$, for the same experiments shown in Fig.~\ref{fig:neck 1}. Each data set is averaged and binned so that the variation in data is indicative of the error, which is much larger at early times when $\kappa_b$ is large. The analytical solution for the Stokes equation and the Darcy solution obtained by the boundary integral simulation are shown for comparison. Additionally, we have included the curvature of the Brownian dynamics simulations from \citet{yue2024coalescing}. There are a few key features illustrated here. The Stokes solution has a larger curvature and increases as $\kappa_b\propto R^2/R_b^3$ when $R_b\rightarrow0$ \cite{paulsen2013approach}. The bridge curvature increases much slower for the Darcy simulation, aside from a sharp, transient drop in the curvature at the initiation of the simulation. At large bridge radii (late times), all experimental data collapse and closely follow the Darcy solution, but are systematically larger. This is presumably due to the presence of an outer fluid in the simulation. Finally, the Brownian dynamics also collapse with the Darcy solution at late times. 

For large $\kappa_b$ (early times), the experimental data plateau since the normalized curvature begins to deviate from the Darcy solution at $\kappa_b R\sim R/h=37.5$. This transition, near $R_b/R = 0.2$, is the same as the transition from $\beta\sim1/2$ to $\beta\sim1/5$ observed in the bridge dynamics, $R_b$ vs. $t$ (Fig.~\ref{fig:neck 1}B). Remarkably, similar behavior has been reported in the particle dynamics simulations from \citet{yue2024coalescing}, where deviations from Darcy's law occur at $R_b/R \sim 0.35$. In those simulations, the crossover is attributed to the finite size of the simulated particles, which limits the local curvature and thus sets the smallest resolvable lengthscale. In contrast, for Hele-Shaw experiments, the relevant length constraint is the gap size $h$ between two plates.
Figure~\ref{fig:neck 1} and Fig.~\ref{fig:contour compare and curvature} clarify the essential role of confinement in determining the coalescence dynamics in a Hele-Shaw cell. At early times, the curvature $1/\kappa_{b}$ is the smallest lengthscale in the system, $1/\kappa_{b} < h < R_b < R$. In this regime, the bridge radius seems to follow $R_b \sim t^{1/2}$. As the bridge and curvature grow in time, eventually $h < 1/\kappa_{b} < R_b < R$, and the gap size $h$ becomes the smallest lengthscale. As a result, Darcy's law is a good approximation of the flow (i.e., Eq.~\ref{velgrad}). 

\section{Jump to contact and the initial regime of coalescence}

The initiation of coalescence between liquid drops can have lasting effects on the subsequent dynamics. For drops in 3D, \citet{sprittles2012coalescence} showed using continuum simulations that the free surface can become ``trapped'' and disappear when two liquid droplets are pressed against each other. \citet{paulsen2013approach} showed experimentally that coalescence initiated when the gap between the liquid interfaces was a few hundred nanometers. More recently, \citet{perumanath2019droplet} used molecular dynamics simulations to illustrate how thermal fluctuations can initiate coalescence, and \citet{deblais2025early} optically and electrically verified this ``jump to contact'' when a drop coalesces with a liquid bath. Finally, \citet{anthony2020initial} showed with continuum simulations that a finite gap between the drops when they coalesce can alter the dynamics, and obfuscate the predictions from theory \cite{eggers1999coalescence}. In these examples, it is often van der Waals forces or other nanoscale phenomena that lead to the initiation of coalescence, which makes the phenomenon difficult to study. 

In our experiments, we also observe a jump to contact, which can be observed in Fig.~\ref{fig:experiment setup}C. By direct imaging, we found that the droplets may ``sense'' each other's existence before the official contact at $t = 0$ s. Figure \ref{fig:meniscus}A shows consecutive frames at the moment of coalescence between two confined drops. The contrast and brightness have been adjusted to highlight the event. The left and middle panels show the approach of the liquid interface as well as a meniscus that connects the vertical outlines. The meniscus appears faint and slightly blurred, yet remains clearly discernible. 
We denote this finite separation between the liquid interfaces as the approach lengthscale $d$, which is measured by fitting both interfaces to circular shapes in the left frame. 

We found that this lengthscale $d$ was nearly proportional to the gap size $h$. Figure ~\ref{fig:meniscus}B shows $d$ vs. $h$, where each symbol represents the mean value obtained from repeated trials. This behavior was consistent over different liquid viscosities and environmental conditions.  \citet{yokota2011dimensional} observed a similar finite distance $d$ during the coalescence of a glycerol drop with a bath in a Hele-Shaw cell. They speculated that the accumulated electric charge and the relative humidity (RH) may determine $d$. Specifically, they reported a larger $d$ in winter compared to summer, presumably due to the differences in humidity. For our experiments, we varied the humidity between 50-70\% and also applied 30 V between the drops, yet observed no discernible change in $d$. We note that silicone oil is a good insulator with lower conductivity  compared to glycerol (which is hygroscopic). 

\begin{figure*}[!]
\centering
\includegraphics[width=0.8\textwidth]{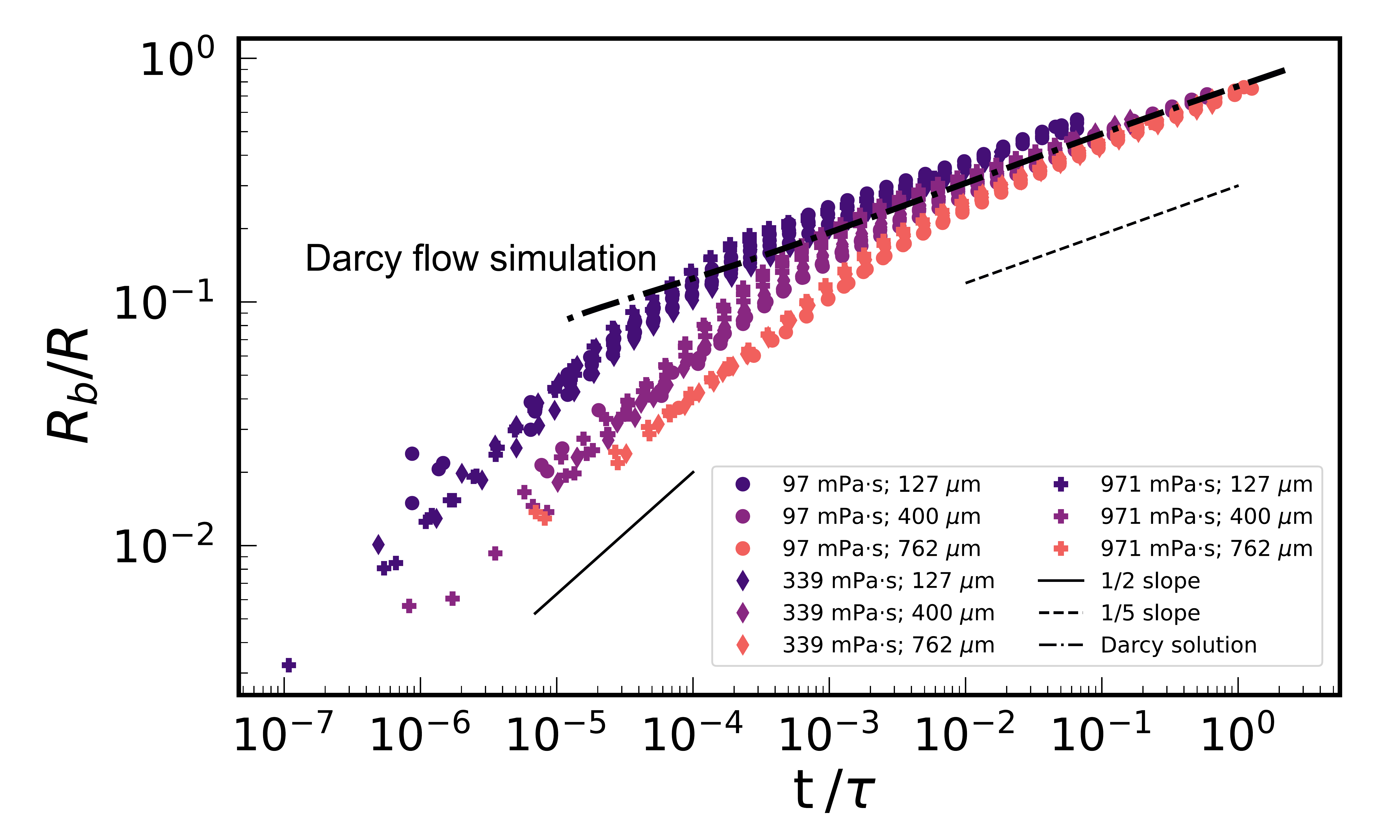}
\caption{Normalized bridge radius as a function of normalized time (Eq.~\ref{taueq}) for various silicone oils ($\eta$ = 97-971 mPa$\cdot$s) with different gap sizes ($h$ = 127-762 $\mu$m), as indicated in the legend. For each combination of $\eta$ and $h$, there are 2-4 datasets shown, indicating the repeatability of the measurements. The Darcy simulation results (dot-dashed line, Eq.~\ref{bintvel}) are overlaid with experimental data for direct comparison. \textcolor{black}{The black dashed line and the black solid line indicate a slope of 1/5 and 1/2, respectively.}}
\label{fig:neck 2}
\end{figure*}

If electrostatic forces are not important, what causes the drops to jump into contact? Figure \ref{fig:meniscus}C illustrates our  proposed mechanism. Silicone oil wets the glass surface of each confining plate, so that its contact angle is 0$^\circ$ and the interface shape between the plates can be approximated by semicircles with radius $h/2$. At first, the two approaching drops remain separated by a thin air layer of thickness $d$. The leading edges of each meniscus are mutually attracted when they begin to contact as a thin wetting film. In Fig.~\ref{fig:meniscus}A, the thick black line represents where the illuminating light is scattered most strongly at the rear of each meniscus where the slope is vertical. Thus, we assume the meniscus extends further from the thick black interface. The attraction and contact between the menisci can be seen in the middle panel of Fig.~\ref{fig:meniscus}A (faintly). Upon contact, the bulk liquid bridge rapidly forms and traps a narrow air pocket within the gap, which eventually disappears and leaves a fully merged liquid bridge. Our attempts to optically image this potential trapped bubble were not successful, so we have not directly verified the hypothetical mechanism in Fig.~\ref{fig:meniscus}C. 


To further illustrate the effects of the gap thickness $h$ on the overall dynamics, Fig.~\ref{fig:neck 2} plots $R_b/R$ vs. $t/\tau$ for three different liquid viscosities and three different values of $h$. At early times, there are dramatic deviations from the Darcy's law predictions. This occurred for two reasons: the gap size directly influences the maximum curvature allowable at the bridge, and also sets the separation between the drops ($d$) when they jump to contact. All data are consistent with a power-law of $\beta \approx $ 1/2 at early times and $\beta \approx$ 1/5 at late times, with a broad transition regime. Since time is normalized by $\tau$ (Eq.~\ref{taueq}), data from different viscosities collapse into groups with different $h$. At late times, the data agree with the Darcy solution, yet shifts upwards (larger prefactor) as $h$ decreases. This is expected since the timescale $\tau$ is derived from Darcy's law, and does not include the limiting effects of $h$ (or $d$) on the initial singularity. 

The next section presents a simple energy balance argument for the $\approx 1/5$ power law in the Darcy regime, which centers on the size of the region where viscous dissipation is strongest. At early times, where $R_b\sim t^{1/2}$ for all data, it is difficult to develop a heuristic energy balance since there are many competing length scales (i.e., $h$, $d$, $R_b$, $\kappa_b$). A more detailed derivation of the early-time dynamics is presented in Sec.~\ref{sec:scaling}. 


\section{Localized Dissipation near the liquid Bridge}

To qualitatively capture the scaling behavior of the bridge growth in the Darcy regime, we can balance the dominant energy input and dissipation in the system that drive coalescence. A similar argument was used in \citet{yokota2011dimensional}, although there are some important differences when choosing the relevant length scales. 
 The rate of change in surface energy is approximated by 
\begin{align}
    \frac{d}{dt} \left( \gamma R_b h \right) = \gamma h \frac{d R_b}{dt}. \label{surface energy}
\end{align}
 Energy dissipation occurs in a volume concentrated around the minimum bridge radius over a lengthscale proportional to the local curvature. Thus, the viscous dissipation is given by
\begin{align}
    \eta \left( \frac{1}{h}\frac{dR_b}{dt}\right)^2 \left(\frac{1}{\kappa_b}\right)^2 h,
    \label{enbaldarcy}
\end{align}
where $h/\kappa_b^2$ is the volume where viscous dissipation is concentrated. We can approximate $\kappa_b$ by $1/\kappa_b \sim R_b^2/R$, as done in deriving Eq.~\ref{deriving 1/5}. The energy balance in the Darcy regime is then given by 
\begin{align}
        \gamma h \frac{d R_b}{dt} = \frac{\eta}{h^2}\frac{R_b^4}{R^2} h \left( \frac{d R_b}{dt} \right)^2. \label{dissipation}
\end{align}
Solving this equation gives:
\begin{align}
    \frac{R_b}{R} \sim \left(\frac{t}{\tau}\right)^{1/5}.
\end{align}
Notably, this energy balance reproduces the same scaling obtained directly from Darcy's law in Eq.~\ref{t 1/5 relation}, providing an independent physical interpretation of the origin of the 1/5 exponent. 

\begin{figure*}[!]
\centering
\includegraphics[width=\textwidth]{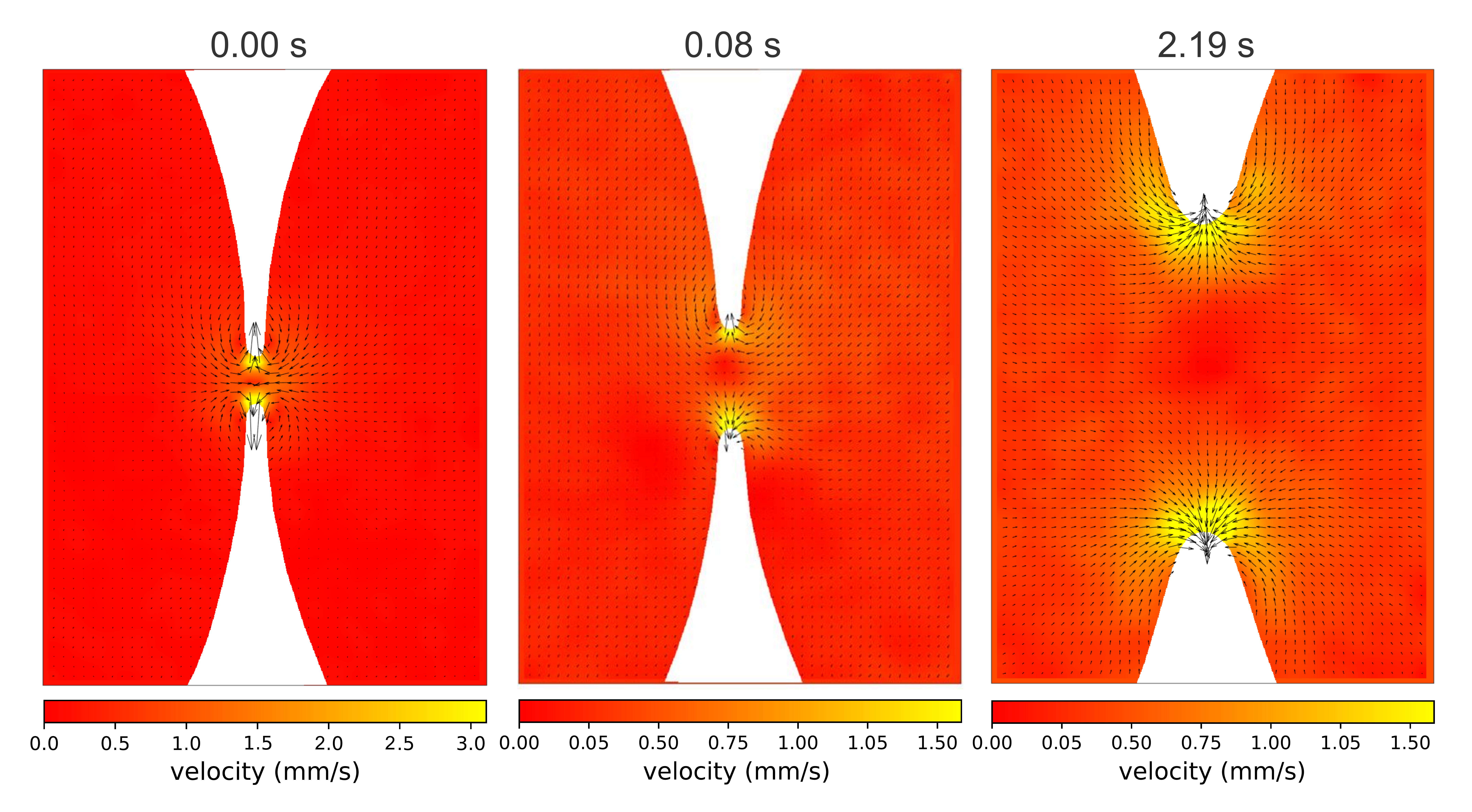}
\caption{Particle Image Velocimetry (PIV) analysis of 339 mPa$\cdot$s silicone oil droplet coalescence using cocoa powder as tracer particles with $h = 400$ $\mu$m. Three snapshots of the flow velocity field measured using the PIVlab application in MATLAB. The white regions represent masks that are excluded from the analysis. Each snapshot uses an independently scaled velocity color map, as indicated by its corresponding color bar, to resolve the spatial structure of the flow at each time.}
\label{fig:piv}
\end{figure*}

The primary difference between the energy balance in \citet{yokota2011dimensional} and ours is the volume of fluid where energy is primarily dissipated. Instead of the volume being localized near the minimum bridge radius, they assume energy is dissipated throughout the bridge, replacing $h/\kappa_b^2$ in Eq.~\ref{enbaldarcy} with $R_b h/\kappa_b$. This results in a scaling exponent $\beta=1/4$, which is hard to distinguish from 1/5 in experimental data. To provide further evidence for the heuristic energy balance argument, we used particle image velocimetry (PIV) to directly measure the localized dissipation.

 

\begin{figure}[!]
\centering
\includegraphics[width=\columnwidth]{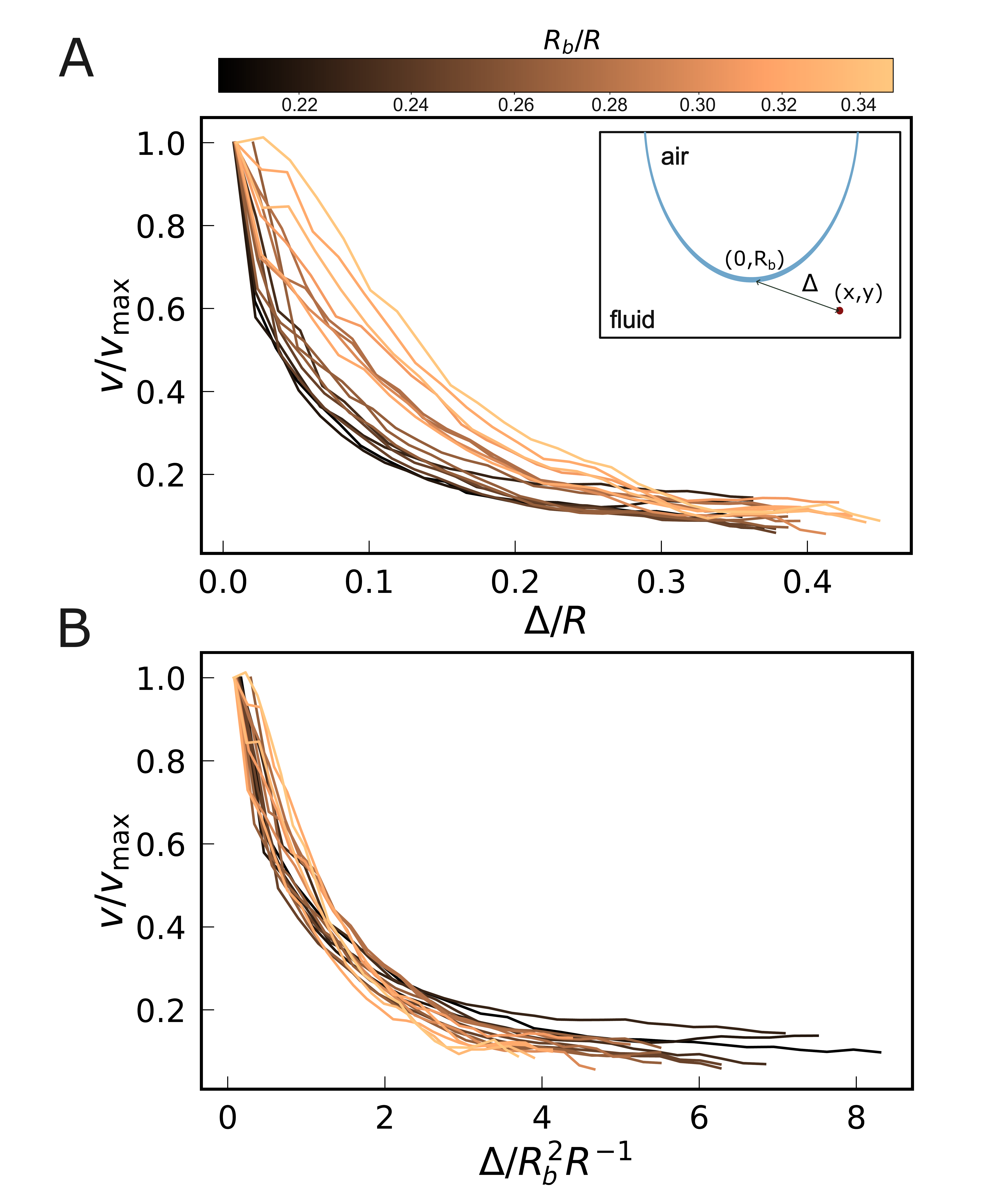}
\caption{Normalized velocity at different $R_b/R$ values as a function of (A) $\Delta/R$ and (B) $\Delta/R_b^2R^{-1}$. The inset shows the cartoon demonstrating all lengthscales involved in this measurement. These data are from the same experiments shown in Fig.~\ref{fig:piv}. }
\label{fig:piv analysis}
\end{figure}

\begin{figure}[!]
\centering
\includegraphics[width=0.6\columnwidth]{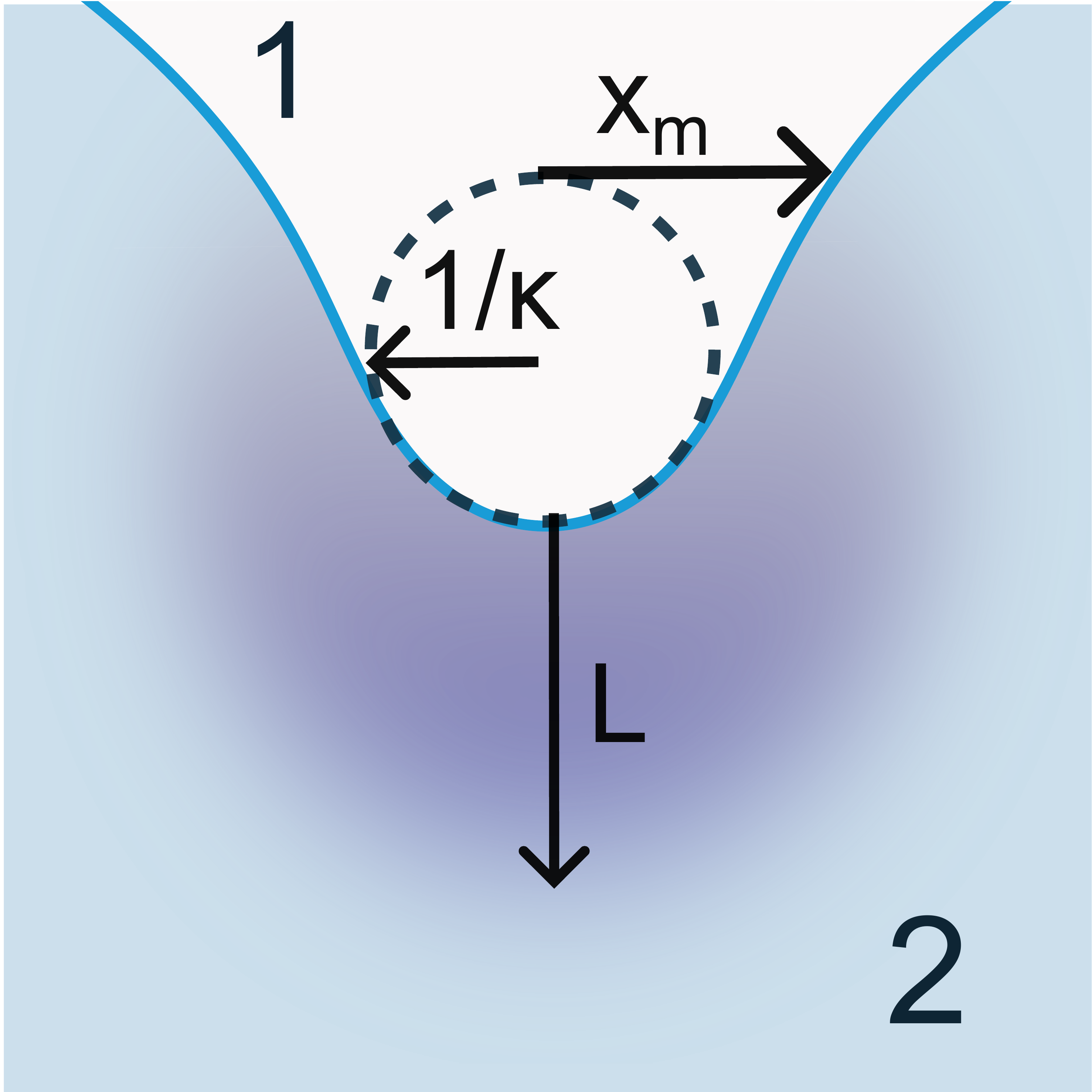}
\caption{A cartoon illustration of the key lengthscales represented in Section VII. The darker part around the curvature represents the gradient of dissipation, where $L$ is the dissipation lengthscale.  }
\label{fig:haicen cartoon}
\end{figure}


Figure \ref{fig:piv} shows that the flow field during coalescence remains strongly localized near the highly-curved bridge region throughout the entire process, while the bulk of the droplet remains nearly stationary. The left panel shows the first frame immediately after contact, which is the early-time regime with $\beta\sim 1/2$. 
At $t=0.08$ s and $t=2.19$ s, the liquid bridge has grown and the velocity decreased, yet the flow remains localized at the bridge interface. This is also true at $t=2.19$ s, when the coalescence dynamics have reached the Darcy regime. Thus, coalescence proceeds like a ``zipper,'' and the bridge advances from flow adjacent to the minimum bridge radius \cite{paulsen2011viscous}. The size of the yellow region in each image indicates the lengthscale over which velocity and pressure magnitudes rapidly decay away from the interface. 

To quantify this decay, we plot the velocity magnitude as a function of the distance to the bridge minimum, defined as $\Delta=\sqrt{x^2+(y-R_b)^2}$, as illustrated in Fig.~\ref{fig:piv analysis}A. A cartoon illustrating how $\Delta$ is defined is shown in the inset. Velocity magnitudes were normalized by the velocity of the minimum bridge radius ($v_{\text{max}}$), and all $\Delta$ were normalized by the drop radius $R$. The velocity clearly decays as $\Delta$ increases, but the lengthscale varies with bridge radius ($R_b/R$). This is obvious from the size of the yellow regions in Fig.~\ref{fig:piv}. We can associate this lengthscale with $L$, the same lengthscale used to derive Eq.~\ref{deriving 1/5} when determining how the pressure decays away from $R_b$. There, we assumed that $L\sim 1/\kappa_b\sim R_b^2/R$. We can test this hypothesis by normalizing $\Delta$ by $R_b^2/R$ instead of just $R$. Figure~\ref{fig:piv analysis} shows that this choice of normalization collapses the data well, and indicates that the bridge width (or inverse curvature) is the defining length scale that controls fluid flow into the singularity.  

Figure~\ref{fig:piv analysis} shows the primary reason to use a localized volume in our energy balance (Eq.~\ref{enbaldarcy}). In the Darcy regime, the flow is localized near the bridge over a lengthscale much smaller than $R_b$. This lengthscale could be $1/\kappa_b$, however, our measurements of the curvature (Fig.~\ref{fig:contour compare and curvature}) are marred by noise due to fitting discrete pixel data. We cannot say with certainty that $1/\kappa_b\sim R_b^2/R$, or if $L\sim 1/\kappa_b$. Next we will show that near $R_b$, the bridge shape is roughly self-similar, with an identifiable, horizontal lengthscale that characterizes the shape evolution. Moreover, we show that analyzing the boundary integral equation for Darcy's law supports this picture. 




\begin{figure*}[t]
\centering
\includegraphics[width=1\textwidth]{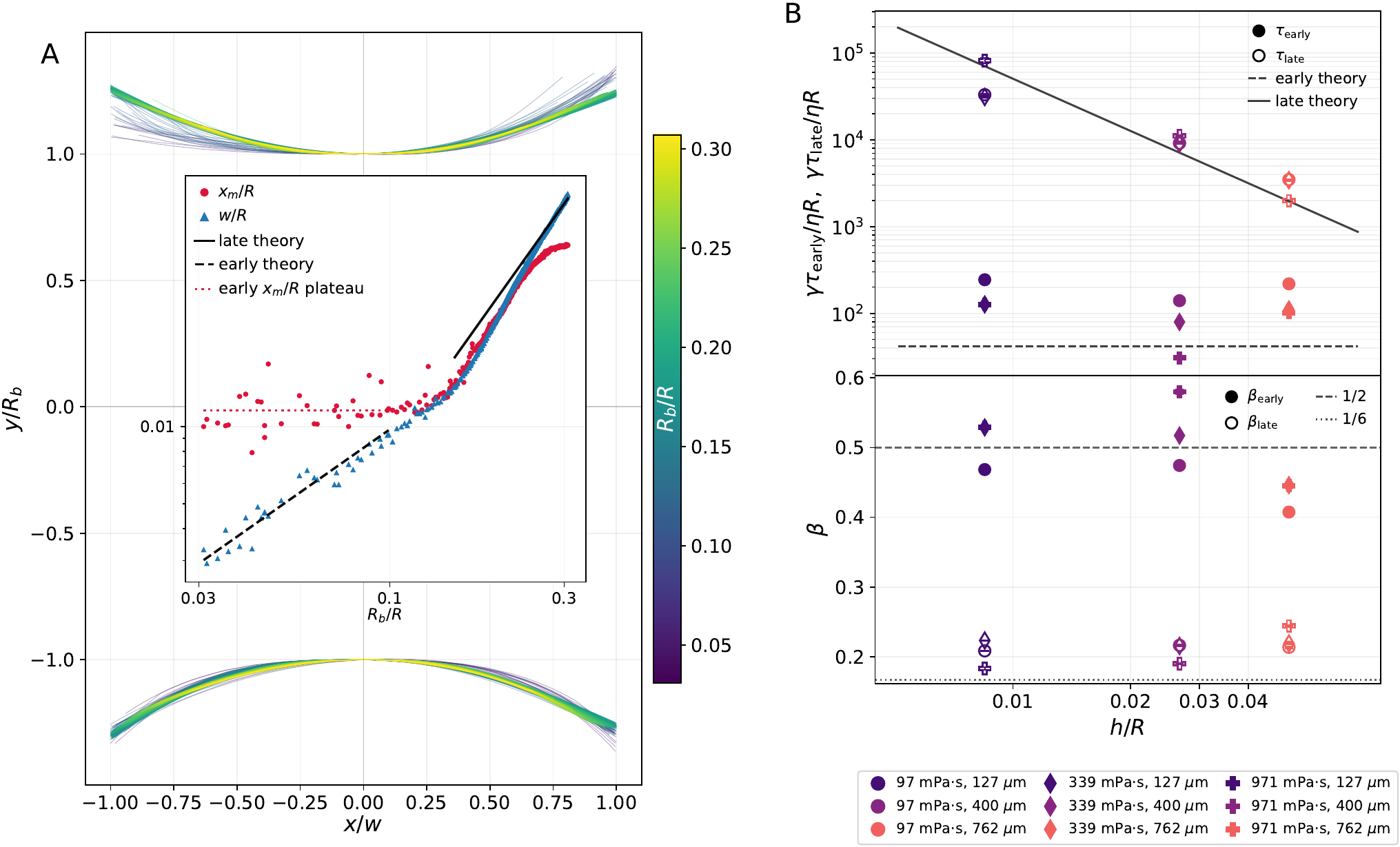}
\caption{Theoretical explanations based on Darcy's law and its Boundary Integral solution. (A) Bridge region profile collapsing based on experimental data with $\eta=971 \text{ mPa}\cdot\text{s}$ and $h = 127\mu \text{m}$. The inset shows the measured $x_m/R$ and $w/R$. The red dotted line is a data fitting line to get the plateau for $x_m/R$ as $d'/(2R) \approx 0.0116$, while the black theoretical lines are directly from Eq.~\ref{eq:xm_Rn} and \ref{eq:early_w} with $H_0$ read from the collapsed profile and $d'$ from the red dotted line directly without further fitting. (B) Early and late stage characteristic time (upper panel) and power law exponent (lower panel) obtained from fitting $R_b(t)/R$ to $(t/\tau)^\beta$. The fitting uses weighted least squares method in log-log scale on data range $R_b/R\leq 0.1$ for early stage and $R_b/R\geq 0.25$ for late stage, with weight as the standard error of mean (SEM) of $R_b(t)$ data from 2-4 experiments for each condition. The error bars give the $95\%$ confidence interval which are too small to be visible. The theoretical lines for $\tau_{\rm{late}}$ and $\tau_{\rm{early}}$ are directly calculated from Eq.~\ref{eq:tau_late} and Eq.~\ref{eq:tau_early} with $H''(0)$ also from the collapsed profile in (A) without further fitting.}
\label{fig:theory}
\end{figure*}

\section{Scaling and Self-similarity}
\label{sec:scaling}

In this section, we will obtain the relation between different length scales relevant to the coalescence as illustrated in Fig.~\ref{fig:haicen cartoon} in a more rigorous way. We start from the energy balance where the decrease of surface energy $E_s$ for the whole contour $\Gamma(t)$ is written as (details in Appendix B):
\begin{equation}
      -\frac{dE_s}{dt} = \gamma h \int_{\Gamma(t)} \kappa(s) v_n(s) ds \approx \frac{32}{15}\gamma h\kappa_b x_m \frac{dR_b}{dt}, \label{eq:E_s}
\end{equation}
where $v_n$ is the normal velocity, and $x_m$ is the length scale representing the matching point of the bridge and the circular region of the drop interface. The main difference between this and our previous energy balance estimation (Eq.~\ref{surface energy}) is that here we do not presume $x_m\sim\kappa_b^{-1}$. 
The viscous dissipation ($dE_v/dt$) can be approximated as (see Appendix \ref{app:enbal}):
\begin{equation}
    \frac{dE_v}{dt} \approx \frac{12\pi\eta}{h} \left(\frac{dR_b}{dt}\right)^2 L^2 \label{eq:E_v}
\end{equation}
where $L$ is the length scale of energy dissipation. Equating Eq.~\ref{eq:E_s} and \ref{eq:E_v}, we can get
\begin{equation}
     \frac{dR_b}{dt} \approx  \frac{8\gamma h^2}{45\pi \eta }\frac{\kappa_b x_m}{L^2}.
     \label{eq:neckgrowth_energy}
\end{equation}

We then start from the Boundary Integral solution of the dimensionless form of Darcy's law (Eq.~\ref{bintvel}). 
For the center of the bridge boundary $\bm{x}_0 = (0,R_b)$, the dimensionless velocity normal to the bridge center can be written as
\begin{equation}
    \tilde 
    v(0) \approx -\frac{1}{2\pi} \int_{-x_m}^{x_m} \frac{\partial \kappa}{\partial x} \frac{x}{x^2+(y-R_b)^2}dx.
\end{equation}
Outside the range $(-x_m,x_m)$ is the circular part where $\kappa$ is almost constant and thus contributes little to the integral. It can be approximated as the point where the curvature changes its sign. Assuming even polynomial expansion for $y-R_b$ and $\kappa(x)-\kappa_b$, we can get (details in Appendix C) leading order approximation for $\tilde v(0)$ as
\begin{equation}
    \tilde v(0) \approx \frac{2\tilde\kappa_b}{\pi \tilde x_m}.
\end{equation}
Thus, the dimensional form bridge growth rate is:
\begin{equation}
    \frac{dR_b}{dt} \approx \frac{R^3}{\tau} \frac{2\kappa_b}{\pi x_m} = \frac{\gamma \kappa_b h^2}{6\pi \eta x_m}, \label{eq:darcy_simple}
\end{equation}
using the relation $v(0)/\tilde v(0) = R/\tau$, $x_m = \tilde x_m R$ and $\kappa_b = \tilde \kappa_b/R$.
Comparing Eq.~\ref{eq:neckgrowth_energy} and \ref{eq:darcy_simple}, we obtain the relation between two length scales $L\sim x_m$. The next step is to get the relation between $x_m$ and $R_b$.

We further assume that the local bridge shape can be described using some self-similar form $y(x,t)=R_b(t)H(X)$ with $X=x/w(t)$. Thus 
\begin{equation}
    \kappa_b =y''(0) = H''(0)R_b/w^2. \label{eq:kappa}
\end{equation}
As the bridge profile should smoothly connect to the outer circle $y_{\text{out}}\approx \sqrt{2Rx}$ at $x\approx x_m$, we have
\begin{align}
    R_b H\left(\frac{x_m}{w}\right) &= \sqrt{2Rx_m} \label{equ:continue}\\  
    \frac{R_b}{w} H'\left(\frac{x_m}{w}\right)  &= \frac{\sqrt{R}}{\sqrt{2x_m}} \label{equ:smooth}
\end{align}
leading to 
\begin{equation}
   \frac{x_m}{R}\approx \frac{H_0^2R_b^2}{2R^2} \label{eq:xm_Rn}
\end{equation}
\begin{equation}
   \frac{w}{R}\approx \frac{H_1H_0R^2_b}{R^2} \label{eq:w_Rn}
\end{equation}
where $H_0\equiv H(x_m/w)$ and $H_1\equiv H'(x_m/w)$. 
Combining the above steps, we can get the relation
\begin{equation}
    L\sim x_m \sim R_b^2/R
\end{equation}
which is consistent with the measurement shown in Fig.~\ref{fig:piv analysis}. Note that the scaling parameter $w$, mathematically, is defined only up to multiplicative constants, and thus needs to be anchored to some physical length scale. Based on Eq.~\ref{eq:xm_Rn} and \ref{eq:w_Rn}, we can choose $w=x_m$ and equivalently, $H_1 = H_0/2$.

Now for the ideal case when both the self-similarity and the connection to a real circle at $x_m$ hold, we can derive the bridge growth scaling law by plugging Eq.~\ref{eq:kappa} and \ref{eq:xm_Rn} into Eq.~\ref{eq:darcy_simple}, which gives
\begin{align}
    &\frac{R_b(t)}{R} \approx \left(\frac{t}{\tau_{\text{late}}}\right)^{1/6} \label{eq:latescaling}\\ 
    &\tau_{\text{late}} = \frac{\pi H_0^6}{96 H''(0)}\tau =\frac{\pi H_0^6 \eta R^3}{8 H''(0)\gamma h^2 } \label{eq:tau_late}.
\end{align}
Note that Eq.~\ref{eq:w_Rn} would lead to a relation $\kappa_bR\sim (R_b/R)^{-3}$ based on Eq.~\ref{eq:kappa}, which contradicts the assumption $x_m\sim \kappa_b^{-1}$ used in the previous rough estimation, giving $\kappa_b R\sim (R_b/R)^{-2}$. The measurement in Fig.~\ref{fig:contour compare and curvature}B gives a power law between $-3$ and $-2$ such that neither assumption (self similarity or $x_m\sim \kappa_b^{-1}$) is a perfect match. The different bridge growth scaling laws for these two assumptions are $t^{1/6}$ and $t^{1/5}$, which are difficult to distinguish using current experimental data (see Fig.~\ref{fig:theory}B for the fitted exponents). This result is not as satisfying as the clear scaling laws derived for Stokes' coalescence problem, but one key point that is confirmed with this work is the localized dissipation for Darcy's coalescence problem, specifically $L\sim x_m$ which depends superlinearly on $R_b$, instead of linearly as assumed in previous studies for coalescing drops confined in a Hele-Shaw cell \cite{yokota2011dimensional}.

In Fig.~\ref{fig:theory}A, we use contours from experiments with $\eta=971 \text{ mPa}\cdot\text{s}$ and $h = 127\mu \text{m}$ to test the self-similarity assumption. This set of data, with the highest $\eta$ and smallest $h$, should be the closest to Darcy's law. During the contour collapsing, $w$ is first anchored to the $x_m$ for a reference frame (the frame closest to $R_b/R=0.2$) to obtain the reference profile $H(X)$. Then, for other frames, $w$ is fitted by minimizing the mean-square difference to this universal profile. Obviously, $w=x_m$ is true for the intermediate regime but not for the early and late stages, due to the gap $d$ before coalescence and the deviation from a circle at $x_m$ during the late stage. The late theory line based on Eq.~\ref{eq:xm_Rn} with $H_0$ directly read from the reference profile shows acceptable consistency with the data considering the several approximated assumptions used. A further check of the late theory is shown in Fig.~\ref{fig:theory}B where $\tau_{\text{late}}\gamma/\eta R$ is plotted over $h/R$ based on Eq.~\ref{eq:tau_late} with both $H_0$ and $H''(0)$ read from the reference profile in Fig.~\ref{fig:theory}A. Consistency is shown not only on the scaling law but also on the prefactor. 



The early stage is influenced by geometric changes induced by the gap $d$ between two drops as the coalescence begins (Fig.~\ref{fig:meniscus}). This is observed in both our Hele-Shaw cell experiments and in Brownian dynamics simulations \cite{yue2024coalescing} where the non-zero $d$ exists (due to different reasons), but not in the numerical solution of Darcy's law without $d$. As a consequence of this gap, $x_m$ and $w$ both deviate from the late stage relations given in Eqs.~\ref{eq:xm_Rn} and \ref{eq:w_Rn}, with $x_m\approx d'/2$ and $w/R\sim R_b/R$. The quantity $d'$ can be obtained by fitting the plateau in Fig.~\ref{fig:theory}A. Then the connecting point of the two stages can be obtained by solving $H_0^2 R_b^2/(2R^2) \approx d'/(2R)$. Plugging this connecting point into the relation $w/R\sim R_b/R$ gives the prefactor, and thus the full relation:
\begin{equation}
    \frac{w}{R} \approx \frac{H_0\sqrt{d'}}{2\sqrt{R}}\frac{R_b}{R} \label{eq:early_w}.
\end{equation}
This relation is plotted as the black dashed line in Fig.~\ref{fig:theory}A without further fitting and its consistency with data confirms that the early stage relation for $x_m$ and $w$ deviate from the late stage relation at the same connecting point. 

Then, plugging $x_m\approx d'/2$ and Eq.~\ref{eq:kappa}, \ref{eq:early_w} into Eq.~\ref{eq:darcy_simple}, we can get
\begin{equation}
    \frac{dR_b}{dt} \approx \frac{4\gamma H''(0)}{3\pi\eta H_0^2} \frac{h^2}{d'^2} \frac{R}{R_b},
\end{equation}
leading to a bridge growth scaling law as:
\begin{align}
    &\frac{R_b(t)}{R} \approx \left(\frac{t}{\tau_{\text{early}}}\right)^{1/2} \label{eq:earlyscaling}\\ 
    &\tau_{\text{early}} = \frac{3\pi H_0^2 \eta R d'^2}{8 H''(0)\gamma h^2 } \label{eq:tau_early}.
\end{align}

Here, $d'$ is fitted from the plateau of $x_m$ in Fig.~\ref{fig:theory}A, which is expected to be related to $d$ measured in Fig.~\ref{fig:meniscus}B, but not necessarily identical. If $d'/d$ is a constant, and assuming $d\sim h$ as measured in Fig.~\ref{fig:meniscus}B, an insensitivity of $\tau_{\text{early}}$ on $h$ is expected. In Fig.~\ref{fig:theory}B, the horizontal dashed line is plotted based on Eq.~\ref{eq:tau_early}. It is worth noting that the consistency between $\tau_{\rm{early}}$'s theory and experiment is not as good as $\tau_{\rm{late}}$ due to several factors. First, the above two mentioned assumptions: $d'\sim d\sim h$ might not be true as it is related to the wetting meniscus on the confining plates and subtle three-dimensional geometry. Second, Darcy's law and the self-similar profile $H(X)$ might not be universal for all experimental conditions used here. It is known that as $\eta$ decreases and $h$ increases, the Hele-Shaw fluid mechanics deviates from the strict assumptions used to derive Darcy's law (see Sec.~\ref{sec:helederiv}). This deviation is significant especially at the early stage when $h\ll R_b$ does not hold. This is evident by the data with the largest gap size, $h$ = 762 $\mu$m, shown in Fig.~\ref{fig:theory}B---the values of $\beta$ deviate significantly from expected at smaller $h$. A mix of Stokes' and Darcy's laws, as well as three- and two-dimensional geometry, contributes to the complexity of the early stage Hele-Shaw coalescence, which influences both the scaling law $\beta$ and characteristic times.

\section{Conclusion and Discussion}

In this study, we investigated the dynamics of two coalescing droplets that are confined in a Hele-Shaw cell. The strong confinement introduces a new class of coalescence dynamics governed by Darcy's law. One key finding of this work is that the temporal evolution of the liquid bridge radius shows two distinct regimes of power-law behavior: $R_b \sim t^{1/2}$ at early times, and $R_b \sim t^{1/5}$ at late times. The initial $R_b \sim t^{1/2}$ regime is a direct consequence of the finite length scale $h$, which is the gap size between the confining plates, and the length scale over which momentum is dissipated. Since flow occurs over scales the size of $h$, Darcy's law is not strictly valid in this regime. In the late-time regime where $R_b \sim t^{1/5}$, the bridge evolution follows Darcy's law directly. A single timescale $\tau$ (Eq.~\ref{taueq}) emerges and captures variations in viscosity, surface tension, and gap size. The 1/5 scaling exponent can be qualitatively derived from Darcy's law, or from a simple energy balance, with some important assumptions about the  width or curvature of the connecting bridge between the drops. 

A second key finding is that the evolution of the bridge radius and curvature are more subtle than simple scaling arguments suggest. The power-law scaling of the bridge radius, $R_b\sim t^\beta$, is sensitive to the horizontal extent of the flow ($w$) surrounding the bridge. This is true in both the early and late-time regimes. We showed that the velocity decays away from the bridge radius over a lengthscale proportional to the bridge width $w$, derived clear relationships between $w$ and $\kappa_b$, yet neither follows a well-defined, power-law scaling over the range of experimental data accessible in the Darcy regime ($R_b/R>0.2$). Thus the 1/5 exponent that is consistent with our data should be taken with caution. Our self-similarity analysis suggests that $1/6<\beta<1/5$, which is challenging to distinguish using our data. Advances in coalescence theory for drops obeying Darcy's law could provide clarity, for example, a full analytical solution similar to Hopper's solution for drops obeying the Stokes equation \cite{hopper1993coalescence1,hopper1993coalescence2}. This is left for future work. 

Another important finding is the ``jump to contact'' between the drops, and its influence on the early-time regime of coalescence dynamics. During coalescence, a singularity is formed from initial conditions, rather than evolving into a self-similar shape, as in drop pinch-off and break up phenomena \cite{eggers2025coalescence, eggers2008physics, eggers1997nonlinear}. Thus, the geometry of the initial interface, nanoscale forces, interstitial fluids, and other complex initial conditions can delay or promote coalescence, and imprint themselves on the early-time dynamics \cite{anthony2020initial,sprittles2014parametric,paulsen2013approach}. For drops in a Hele-Shaw cell, we showed that the menisci between the drops interact well before the sharp, visible boundaries come into contact. Understanding and controlling the initiation of coalescence singularities is imperative in numerous industrial and natural applications, from emulsion coarsening to raindrop collisions.


More broadly, we emphasize that coalescing drops and fluids obeying Darcy's law are not limited to Newtonian liquids confined between solid boundaries. Recently,  \citet{yue2024coalescing} showed that particles obeying Brownian dynamics can form clusters whose hydrodynamics are described by Darcy's law. This is because the particles are coupled with a dissipative background. The authors also showed that coalescence data from other living systems, from cell nucleoli to bacterial colonies, may all obey Darcy's law. Therefore, the experiments and analysis presented here provide a unified physical picture for coalescence in many confined systems where transport is not solely governed by a single fluid viscosity, but by the complexities of the external environment.

\begin{acknowledgements}
We acknowledge Daniel Sussman for many insightful conversations. J.W., J.C.B., N.V., and T.W. were supported by  
the Gordon and Betty Moore
Foundation, grant DOI 10.37807/gbmf12256. H.Y. was supported by the National Science Foundation grant number 2440375. 
\end{acknowledgements}

\bibliographystyle{apsrev4-2}
\bibliography{ref}

\appendix

\section{Boundary integral formulation}
\label{appendA}

Hopper's solution is derived from incompressible Stokes flow \cite{hopper1993coalescence1,hopper1993coalescence2}:
\begin{align}
\bm{\nabla}P=\eta\bm{\nabla}^2{\bm{v}}\\
\bm{\nabla}\cdot{\bm v}=0.
\end{align}
Here, $\eta$ is the fluid viscosity. In contrast, for fluids that obey Darcy's law, the resulting equations are: 
\begin{align}
\bm{\nabla}P&=-\alpha{\bm v}\\
\bm{\nabla}\cdot{\bm v}&=0,
\end{align}
where $\alpha$ is the inverse mobility. Due to incompressibility, both equations reduce to solving Laplace's equation for the pressure:
\begin{align}
\bm{\nabla}^2P&=0\\
P_1 - P_2&=\gamma\kappa.
\end{align}
The difference between solving these problems is how the velocity is obtained once the pressure field is known.

Consider two disks of Darcy fluid coalescing in a background Darcy fluid. The equation for each fluid is:
\begin{align}
\bm{\nabla}P_1&=-\alpha_1{\bm v}_1,\\
\bm{\nabla}P_2&=-\alpha_2{\bm v}_2.
\end{align}
We can make the equations dimensionless using the radius of a single disk $R$, a time scale $T=\alpha_1R^3/\gamma$, and a mass scale $M=\gamma T^2$. The equations then become:
\begin{align}
\bm{\nabla}P_1&=-{\bm v}_1,\\
\bm{\nabla}P_2&=-K{\bm v}_2,
\end{align}
where $K=\alpha_2/\alpha_1$, and all quantities are now dimensionless. The limit $K\rightarrow 0$ corresponds to a high viscosity interior fluid, and a low viscosity outer fluid where the pressure is nearly uniform. We define the velocity potentials:
\begin{align}
\phi_1&=-P_1,\\
\phi_2&=-P_2/K,\\
q&=\phi_2-\phi_1.
\end{align}
The normal derivative of $\phi$ is now continuous using a double-layer potential (the normal component of the velocity is continuous). To derive a boundary integral equation, we start with the boundary condition:
\begin{align}
    P_1-P_2&=\kappa,\\
    K\phi_2-\phi_1&=\kappa.
\end{align}
where $\kappa$ is now the dimensionless mean curvature. The second equation is found just by plugging in the definition of $\phi$ into the first. Solving for $\phi_1$, we get
\begin{equation}
    \phi_1=\dfrac{\kappa-q K}{K-1}.\label{phiq}
\end{equation}

Now, using the double layer formulation from Pozrikidis \cite{pozrikidis1992boundary}:
\begin{equation}
q({\bm x}_0)=2\oint\bm{\nabla} G({\bm x}_0,{\bm x})\cdot \bm n({\bm x})q({\bm x})ds-2\phi_1({\bm x}_0)
\end{equation}
This is the equation assuming we know what $\phi_1$ is. We define the boundary integral as $I(q)$, so that
\begin{equation}
q({\bm x}_0)=2I(q)-2\phi_1({\bm x_0}).
\end{equation}
We plug in Eq.~\ref{phiq} into the integral equation, and rearrange:
\begin{equation}
q=2\dfrac{1-K}{K+1}I(q)+\dfrac{2}{K+1}\kappa.
\end{equation}
This is the integral equation for $q$. For a given boundary shape, we can compute the curvature $\kappa$, then solve the integral equation for $q$. Once $q$ is known, we then use it to compute the normal velocity from a vector potential:
\begin{equation}
    {{\bm v}}(x_0) = \oint \bm{\nabla} G({\bm x}_0,{\bm x})\times\left[\bm{\nabla}q({\bm x}) \times {\bm n}({\bm x})\right] ds(\bm x). \label{eq:bim}
\end{equation}

If $K=1$, the inner and outer fluid have equal mobilities, and there is no need to solve an integral equation since $q=\kappa$. To obtain the $K=1$ solution shown in Fig.~\ref{fig:neck 1}B, we used custom MATLAB 2020b code starting from the initial shape $r(\theta)=2(a-\sin^2{\theta})^{1/2}$ with $a=1.001$, which is a polar-coordinate representation of Hopper's solution for the Stokes equation \cite{hopper1993coalescence1} at a very early time. 1200 points are used to represent the interface. The interface was advanced using an adaptive step size fifth-order Dormand-Prince method \cite{press2007numerical}. The initial step size was $dt=10^{-12}$, and was adaptive based on the truncation error of the method. At each step, normal velocity of the 1200 points is calculated using Eq.~\ref{eq:bim} to update their position. After each update, we redistributed the 1200 points based on the arc-length $ds$ between neighboring marker points using cubic spline interpolation with end conditions (\emph{csape} function in MATLAB). We required a relation $ds \propto (x^2+y^2)^k$ to ensure a denser distribution of points near the bridge region where the curvature is the highest.

\section{Energy balance}
\label{app:enbal}

We define the interface of the coalescing drops to be a parametrized curve $\bm{r}(\theta, t)$, and thus the total length is 
\begin{equation}
    \Gamma(t) = \oint ds = \oint |\partial_\theta\bm{r}|d\theta.
\end{equation}
The time derivative of total interface length is
\begin{equation}
    \frac{d\Gamma}{dt} = \oint \frac{\partial_\theta \bm{r}}{|\partial_\theta \bm{r}|}\cdot \partial_t\partial_\theta \bm{r} d\theta = \oint \bm{t}\cdot \partial_\theta \partial_t \bm{r} d\theta,
\end{equation}
where $\bm{t}$ is the tangent unit vector and the second step uses the relation $\partial_t \partial_\theta \bm{r} = \partial_\theta \partial_t \bm{r}$.
We can decompose the interface velocity into tangent and normal components
\begin{equation}
    \partial_t \bm{r} = v_t \bm{t} + v_n \bm{n},
\end{equation}
and find that
\begin{equation}
    \bm{t}\cdot \partial_\theta \partial_t \bm{r} = \partial_\theta v_t -v_n \kappa |\partial_\theta \bm{r}|,
\end{equation}
where we used the relation $\partial_\theta \bm{n} = -\kappa |\partial_\theta \bm{r}|\cdot \bm{t}$.
Because $\oint \partial_\theta v_t d\theta=0$, we find that
\begin{equation}
    \frac{d\Gamma}{dt} = -\oint v_n\kappa |\partial_\theta \bm{r}|d\theta = -\oint v_n\kappa ds.
\end{equation}
We now assume that in the circular part, $v_n\approx 0$, so the integral is limited to the bridge region:
\begin{equation}
    \frac{d\Gamma}{dt}\approx -2\int_{-x_m(t)}^{x_m(t)} v_n \kappa ds \approx -4C_s x_m(t) \frac{dR_b}{dt}\kappa_b.
\end{equation}
The constant $C_s$ is a geometric prefactor smaller than $1$ because both $v_n$ and $\kappa$ decrease to roughly zero from the minimum bridge radius, corresponding to  $v_n(0)= dR_b/dt$ and $\kappa(0)=\kappa_b$. Considering the symmetry about $x=0$, the lowest order approximations for $\kappa(x)$ and $v_n(x)$ are
\begin{align}
    \kappa(x) &= \kappa(0) (1-x^2/x_m^2)\\
    v_n(x) &= v_n(0) (1-x^2/x_m^2).
\end{align}
Then, the geometric prefactor $C_s=\int_0^1 (1-X^2)^2dX = 16/30$. Thus, the rate of surface energy decrease is
\begin{equation}
    -\frac{dE_s}{dt} = \frac{32}{15}\gamma h\kappa_b x_m \frac{dR_b}{dt} .
\end{equation}

For the viscous dissipation, we assume the velocity gradient is dominated by the parabolic velocity profile in the $z$-direction (Eq.~\ref{darcy der: v}). The rate of viscous dissipation is then approximated by: 
\begin{equation}
    \frac{dE_v}{dt}=12\eta\int  \frac{|\bm{v}(x,y)|^2}{h}dx dy.
\end{equation}  
Considering the in-plane decay of speed from $dR_b/dt$ at the bridge tip $(0, \pm R_b)$ to approximately zero at distance $L$ away (Fig.~\ref{fig:haicen cartoon}), the above equation can be written as
\begin{equation}
    \frac{dE_v}{dt}\approx \frac{12\eta C_v}{h} \left(\frac{dR_b}{dt}\right)^2 L^2
\end{equation}
where $C_v$ is another geometric factor. 
If we assume an exponential decay based on the shape in Fig.~\ref{fig:piv analysis}, 
\begin{equation}
    |\bm{v}(x,y)| = \frac{dR_b}{dt}\exp\left(-\frac{\Delta}{L}\right),
\end{equation}
where $\Delta = \sqrt{x^2+(y-R_b)^2}$, 
the total viscous dissipation is then
\begin{align}
    \frac{dE_v}{dt} &= 4\pi \frac{12\eta}{h}\int_0^\infty d\Delta \left(\frac{dR_b}{dt}\right)^2 \exp\left(-\frac{2\Delta}{L}\right)\Delta  \\
    &= \frac{12\pi\eta}{h} \left(\frac{dR_b}{dt}\right)^2 L^2,\nonumber
\end{align}
where the factor $4\pi$ comes from the angular part of the integral from the upper and lower branches of the bridge, each contributing approximately $2\pi$. Of course, the exact angular integration range is not the full $2\pi$ and decreases as $R_b$ increases. 
Equating the surface energy decreasing rate to the viscous dissipation rate, we have
\begin{equation}
    \frac{dR_b}{dt} = \frac{8\gamma h^2}{45\pi \eta }\frac{\kappa_b x_m}{L^2}
\end{equation}


\section{Leading order dynamics}
We can derive the leading order dynamics of the bridge by starting from the boundary integral solution of Darcy's law (Eq.~\ref{bintvel}):
\begin{equation}
    \bm{v}(\bm x_0) = \frac{\gamma h^2}{12\eta}\oint_D \nabla G(\bm{x}_0,\bm{x})\times[\, \nabla \kappa(\bm{x}) \times  \bm{n}(\bm{x})\,] ds(\bm{x}).
\end{equation}
Here we have used the dimensional version of the equation. The Green's function for the Laplace equation in 2D is given by $G(\bm{x}_0,\bm{x}) = -\frac{1}{2\pi} \ln |\bm{x}_0 - \bm{x}|$. In our two-dimensional geometry and for the center of the bridge boundary, $\bm{x}_0 = (0,R_b)$, the velocity normal to the bridge center can be written as
\begin{align}
    v(0) &\approx -\frac{\gamma h^2}{24\pi \eta} \int_{-x_m}^{x_m} \frac{\partial \kappa}{\partial x} \frac{x}{x^2+(y-R_b)^2}dx\
    \label{eq:ACvint}
\end{align}
Here, $(-x_m,x_m)$ is the range of the bridge region outside of which the curvature $\kappa$ is almost constant and thus contributes little to the integral. It can be approximated as the point where the curvature changes its sign. 

Considering even symmetry, the polynomial expansions of $y-R_b$ and $\kappa(x)-\kappa_b$ are
\begin{align}
    y(x)-R_b &= a_2 x^2+a_4x^4+\cdots, \\
    \kappa(x) - \kappa_b &= \kappa_2 x^2 + \kappa_4 x^4 + \cdots.
\end{align}
We note these coefficients are related because $\kappa(x) = \frac{y''(x)}{(1+y'(x)^2)^{3/2}}$. Thus, the integral in Eq.~\ref{eq:ACvint} becomes
\begin{equation}
    I = 4\int_0^{x_m}  \frac{(\kappa_2 + 2\kappa_4 x^2 + \cdots)}{1 + (a_2 x +a_4 x^3+\cdots)^2}dx. \label{seq:v_integral}
\end{equation}
Using the approximation for small $x$:
\begin{align}
    \frac{1}{1+(a_2 x +a_4 x^3+\cdots)^2}\approx 1- a_2^2 x^2
\end{align}
The integrand in Eq.~\ref{seq:v_integral} becomes
\begin{equation}
    \kappa_2 + \left(2\kappa_4 - \kappa_2a_2^2\right)x^2 
\end{equation}
So, the leading order approximation of the integral is 
\begin{equation}
    I \approx 4\kappa_2 x_m. 
\end{equation}
 Since $\kappa(x_m) = 0$, we have
\begin{equation}
    \kappa_b + \kappa_2 x_m^2 \approx 0.
\end{equation}
Thus, the dynamics of the bridge growth become
\begin{equation}
    v(0)=\frac{dR_b}{dt} = \frac{\gamma h^2\kappa_b}{6\pi\eta x_m}.
\end{equation}

\end{document}